\documentclass[twocolumn,twoside,superscriptaddress,showpacs,floatfix]{revtex4-2}
\usepackage{amssymb,amsmath,graphicx,multirow,natbib,hyperref,color}
\hypersetup{colorlinks,citecolor=black,filecolor=black,linkcolor=black,urlcolor=black}

\begin{document}
\title[]{Improved Numerical Scheme for the Generalized Kuramoto Model}
\author{Hyun Keun \surname{Lee}} 
\affiliation{Department of Physics, Sungkyunkwan University, Suwon 16419, Korea}
\author{Hyunsuk \surname{Hong}}
\email{Corresponding author: hhong@jbnu.ac.kr}
\affiliation{Department of Physics and Research Institute of Physics and Chemistry, Jeonbuk National University, Jeonju 54896, Korea}
\author{Joonhyun \surname{Yeo}}
\affiliation{Department of Physics, Konkuk University, Seoul 05029, Korea}

\date{\today}
\begin{abstract} 
We present an improved and more accurate numerical scheme for a generalization of the Kuramoto model of coupled phase oscillators to the three-dimensional space. The present numerical scheme relies crucially on our observation that the generalized Kuramoto model corresponds to particles on the unit sphere undergoing rigid body rotations with position-dependent angular velocities. We demonstrate that our improved scheme is able to reproduce known analytic results and capture the expected behavior of the three-dimensional oscillators in various cases. On the other hand, we find that the conventional numerical method, which amounts to a direct numerical integration with the constraint that forces the particles to be on the unit sphere at each time step, may result in inaccurate and misleading behavior especially in the long time limit. We analyze in detail the origin of the discrepancy between the two methods and present the effectiveness of our method in studying the limit cycle of the Kuramoto oscillators.
\end{abstract}



\maketitle

\section{\label{intro}Introduction}
The Kuramoto model~\cite{Kuramoto7584} is a widely-used mathematical model for 
studying the synchronization behavior in populations of coupled 
oscillators~\cite{Strogatz00,Pikovsky03,Acebron05}. 
The model consists of coupling 
terms describing the interactions among phase oscillators, and 
natural frequencies randomly distributed according to a distribution function.
When the coupling strength surpasses the diversity of natural frequencies, collective synchronization emerges.
Variations of the Kuramoto model have been proposed, including time delays~\cite{vkm1}, 
inertia effects~\cite{vkm2-1, vkm2-2}, and thermal noise~\cite{vkm3}.
Coupled oscillator models have also been used 
to study the swarming~\cite{tswm,Kevin17,Kevin19,Lee21,tswm2,tswm3} and flocking~\cite{tflk,tflk2} behaviors of 
natural and artificial systems.

One of the recent interests in the study of Kuramoto models is to 
increase the dimension of the phase variable or the embedding space.
The pioneering work in this direction is the 
{\textit{consensus}} model, where an individual's opinion is represented by a vector~\cite{reza2006}. 
Later, it was shown that the {\textit{algebraic}}
generalization can be done systematically~\cite{lohe2009} 
using the rotation group theory~\cite{sakurai,tung,arfken}.  
With these tools, the 
Kuramoto model in higher dimensions has been steadily 
studied~\cite{zhu2013,tanaka2014,lohe2,lohe3,Ott2019,fplk,fplk1.5,
fplk2,fplk3,lipton2021}. These studies often illustrate the results 
in figures only in three dimensions as 
the visualization in higher dimensions is not practical.
The geometric intuition in three dimensions plays an 
irreplaceable role in understanding 
the behavior of the Kuramoto oscillators, and therefore 
a solid description of the model in three dimensions 
is important.

In this work, we focus on the three-dimensional generalization of the Kuramoto model, 
and develop a new and improved method for the numerical integration
of the model. In three dimensions, the 
Kuramoto oscillators are represented by particles moving on the unit sphere.
As we will explain in detail later, the present numerical scheme is motivated by
the observation that the generalized Kuramoto model can be rewritten as 
a vector product form, which indicates that the particles undergo
rigid body rotations on the unit sphere with position-dependent angular velocity vectors.
This is in contrast to the 
conventional method~\cite{reza2006}, which is just a straightforward numerical integration
with the constraint enforcing the particles on the unit sphere at each time step. 
Although it is not stated explicitly in literatures, 
we believe that the conventional method has been regarded as a standard choice 
of the numerical integration scheme in many other works
(we will provide some evidence for this in Sec.~\ref{summary}).
In this paper, we show that the conventional method contains a numerical artifact,
which may result in misleading behavior of the Kuramoto oscillators.
We show that this problem is
caused by the particular use of the constraint producing numerical artifacts.
On the other hand, we demonstrate below that our improved method successfully reproduces
available analytic solutions as well as the behaviors of the oscillators which are expected 
on physical grounds.

This paper is organized as follows. In Sec.~\ref{mdl}, we present our 
extension of the Kuramoto model to three dimensions in a vector product form. 
The new and improved numerical method follows directly 
from the model equation and is explained in Sec.~\ref{sn:ni} along with 
conventional one. We also present an analytic argument on the numerical artifact of the conventional method. 
In Sec.~\ref{sn:cd}, 
we present numerical results obtained by our method and show that they are consistent 
with analytic solutions and expected behaviors of the oscillator. We also
compare them to those produced by the conventional 
numerical method and point out how it can lead to misleading behaviors.  
This paper concludes with the brief summary and perspective 
in Sec.~\ref{summary}.

\section{Kuramoto model in vector product form}
\label{mdl}
The Kuramoto model is given by a differential equation for the dynamics of coupled phase oscillators:
\begin{equation}
\dot\theta_i = \omega_i + \frac{K}{N}\sum_{j=1}^N \sin(\theta_j-\theta_i)\,,
\label{eq:K}
\end{equation}
where $\theta_i \in [0,2\pi)$ is the phase of the oscillator $i$ and $\omega_i$ is its natural frequency.  The $K > 0$ is the coupling constant and $N$ is the number of oscillators in the system. Generalizations of the Kuramoto model to higher dimensions have been considered in
Refs.~\cite{reza2006,lohe2009,zhu2013}. Below we present an alternative construction of the generalized Kuramoto model in
three dimensions, whose form will play an important role in developing a new numerical scheme.

We consider a circle of unit radius at the origin of the $x$-$y$ plane in the three-dimensional space. Then, in the Cartesian coordinate system, we introduce the position vector for oscillator $i$ of phase $\theta_i$ on the circle: ${\bf{r}}_i =(x_i,y_i,0)=(\cos{\theta_i},\sin{\theta_i},0)$. This way, the oscillator of phase $\theta_i$ can be also identified with a particle or an agent at position ${\bf{r}}_i$. Below, for convenience, we interchangeably use the term,
oscillator, particle, or agent. Similarly, for another oscillator $j$ on the unit circle, it reads that ${\bf{r}}_j=(\cos{\theta_j},\sin{\theta_j},0)$. For ${\bf{r}}_i$ and ${\bf{r}}_j$ written this way, one can observe that i) $\dot {\bf{r}}_i=(-\sin\theta_i,\cos\theta_i,0)\dot\theta_i$, ii) 
${\bf w}_i \times {\bf{r}}_i =(-\sin\theta_i,\cos\theta_i,0)\omega_i$ for ${\bf w}_i=(0,0,\omega_i)$ with the 
vector product operator $\times$~\cite{arfken}, and iii) $({\bf{r}}_i\times{\bf{r}}_j)\times{\bf{r}}_i =(-\sin\theta_i,\cos\theta_i,0)\sin(\theta_j-\theta_i)$.

Using these observations, one can rewrite Eq.~\eqref{eq:K} as
\begin{equation}
\dot {\bf{r}}_i = {\bf w}_i\times {\bf{r}}_i
	+\frac{K}{N}\sum_{j=1}^N 
	({\bf{r}}_i\times {\bf{r}}_j)\times {\bf{r}}_i\,. 
\label{eq:2K}
\end{equation}
This formula is straightforwardly obtained by
applying $(-\sin\theta_i,\cos\theta_i,0)$ to both sides of Eq.~\eqref{eq:K}. 
We then introduce ${\bf k} \equiv K\sum_j {\bf{r}}_j/N = K {\bf{r}}_{\rm CM}$, where ${\bf r}_{\rm CM}$ is the position center of the agents. 
Then, Eq.~\eqref{eq:2K} is simply rewritten as
\begin{equation}
\dot {\bf{r}}_i = ({\bf{w}}_i-{\bf{k}}\times 
	{\bf{r}}_i)\times {\bf{r}}_i
	\equiv {\bf v}_i \times {\bf{r}}_i \,.
\label{eq:2Kf}
\end{equation} 
The minus sign on the right hand side of Eq.~\eqref{eq:2Kf} comes from ${\bf{r}}_i \times {\bf{r}}_j = -{\bf{r}}_j\times {\bf{r}}_i$.

Equation~\eqref{eq:2Kf} is the spatial-coordinate 
representation of the standard Kuramoto model in Eq.~\eqref{eq:K} 
when ${\bf r}_i=(\cos\theta_i,\sin\theta_i,0)$ and ${\bf w}_i=(0,0,\omega_i)$. 
The generalized Kuramoto model we consider in this paper
is Eq.~\eqref{eq:2Kf} obtained after the restrictions
on ${\bf r}_i$ and the angular velocity ${\bf w}_i$ are removed.
This means that ${\bf r}_i$ is now allowed to reside on the unit sphere 
centered at the origin not just on the $x$-$y$ plane and  
${\bf w}_i$ can point in a general direction not just the $z$-direction.

Since the vector product $\times$ is a linear operation, 
one may consider a matrix $\Omega_{{\bf w}_i}$ that satisfies 
$\Omega_{{\bf w}_i} {\bf r}_i = {\bf w}_i\times {\bf r}_i$. 
From the detail of the operation $\times$, it is given by
\begin{equation}
\Omega_{{\bf w}_i} = 
	\begin{pmatrix}
          0 & -\omega_{i,z} & \omega_{i,y} \\
	  \omega_{i,z} & 0 & -\omega_{i,x} \\
	  -\omega_{i,y} & \omega_{i,x} & 0  
	\end{pmatrix}
\label{eq:Om}
\end{equation}
for ${\bf{w}}_i=(\omega_{i,x},\omega_{i,y},\omega_{i,z})$.
Additionally, by the identity ${\bf{a}}\times ({\bf{b}}\times {\bf{c}}) = ({\bf{a}}\cdot {\bf{c}}){\bf{b}}-({\bf{a}}\cdot {\bf{b}}){\bf{c}}$~\cite{arfken}, Eq.~\eqref{eq:2Kf} is rewritten as
\begin{equation}
\dot {\bf{r}}_i 
=\Omega_{{\bf w}_i} {\bf{r}}_i +{\bf{k}} - ({\bf{k}}\cdot {\bf{r}}_i){\bf{r}}_i\,.
\label{eq:2Kc}
\end{equation}
We remark that Eq.~\eqref{eq:2Kc} is equivalent to 
the generalized Kuramoto equations 
studied in Ref.~\cite{reza2006,lohe2009,zhu2013}.
It is known that Eq.~\eqref{eq:2Kc} is still valid in
higher dimensions above three with a skew-symmetric matrix~\cite{arfken} 
$\Omega_{{\bf w}_i}$ given through ${\bf w}_i$.
When we refer to Eq.~\eqref{eq:2Kc} in this work, 
it means the three-dimensional case.

\section{\label{sn:ni} Numerical integration methods}

Although Eq.~\eqref{eq:2Kf} is identical to Eq.~\eqref{eq:2Kc}, 
it provides a key insight for a numerical integration scheme,
which is not obvious if the equation is written in the form of Eq.~\eqref{eq:2Kc}.
Equation \eqref{eq:2Kf} suggests that the particles perform rigid body rotations
with appropriate angular velocities. In the following subsection, we develop 
a numerical integration scheme based on this observation. After that, we analyze the 
conventional numerical method and show how this method can lead
to numerical artifacts.

\subsection{\label{sn:nir} Improved numerical method: Evolution by a direct rotation}

The differentiable motion on a sphere is locally an arc with its own
curvature. Thus specifying the associated circle or, equivalently, 
the rotation axis is most important in describing the motion 
on a sphere. According to the physics of   
rigid body rotation~\cite{goldstein}, 
the parenthesis part of Eq.~\eqref{eq:2Kf}, 
${\bf v}_i = {\bf v}_i(t) = {\bf w}_i-{\bf k}(t)\times{\bf r}_i(t)$, 
is parallel to the axis of rotation of ${\bf r}_i(t)$ and,
furthermore, exactly corresponds to  
the angular velocity of the rotation.
When this velocity is considered constant over a small time mesh $\Delta t$, 
the position update ${\bf r}_i^{\rm h}(t+\Delta t)$ is given by
\begin{equation}
\label{eq:hr}
	{\bf r}_i^{\rm h}(t+\Delta t)
	=R\left({\bf v}_i(t)\Delta t\right){\bf r}_i(t)\,,
\end{equation}
where $R({\bf a})$ is the rotation matrix by 3-dimensional angle vector $\bf a$.
For ${\bf a}
=\lVert{\bf a}\rVert(\sin\phi\cos\theta,\sin\phi\sin\theta,\cos\phi)$, 
it is given~\cite{arfken} that
\begin{equation}
\label{eR}
R({\bf a})=R_z(\theta)R_y(\phi)R_z(\lVert{\bf a}\rVert)R_y(-\phi)R_z(-\theta)\,,
\end{equation}
where
\begin{equation}
\label{Rz}
R_z(\alpha)=
\begin{pmatrix}
\cos\alpha & -\sin\alpha & 0 \\
\sin\alpha & \cos\alpha & 0 \\
0 & 0 & 1 
\end{pmatrix}
\end{equation}
and
\begin{equation}
\label{Ry}
R_y(\alpha)=
\begin{pmatrix}
\cos\alpha & 0 & \sin\alpha \\
0 & 1 & 0 \\
-\sin\alpha & 0 & \cos\alpha 
\end{pmatrix}\,.
\end{equation}
The numerical integration method we propose in this paper
is Eq.~\eqref{eq:hr} with the rotation matrix given
by the expressions in Eqs.~\eqref{eR}, \eqref{Rz}, and \eqref{Ry}.

Note that the new scheme, Eq,~\eqref{eq:hr} is possible only after we write the three-dimensional 
Kuramoto model in the form of Eq.~\eqref{eq:2Kf}.
The rotation axis in Eq.~\eqref{eq:hr}
changes in a discrete way per $\Delta t$,
and its time series approaches
the profile of the continuously varying rotation axis 
in Eq.~\eqref{eq:2Kf}, as $\Delta t$ decreases. 
This way, the rate of the difference equation in Eq.~\eqref{eq:hr}  
converges 
the differential equation Eq.~\eqref{eq:2Kf} 
(or, equivalently, Eq.~\eqref{eq:2Kc} in three dimensions) 
in the $\Delta t\to 0$ limit.
We therefore expect that a numerical integration by Eq.~\eqref{eq:hr} 
yields proper results with small $\Delta t$.

\subsection{\label{sn:cim} Conventional method and its artifacts}

The conventional method for the numerical integration of 
differential equation 
Eq.~\eqref{eq:2Kc}
begins with considering ${\bf r}_i(t)+\dot{\bf r}_i(t)\Delta t$ 
for small 
$\Delta t$, and then enforcing the normalization that makes the agents stay 
on the unit sphere. Thus when the position at later time is denoted by 
${\bf r}_i^{\rm c}(t+\Delta t)$, we have
\begin{equation}
	\label{eq:cni}
{\bf r}_i^{\rm c}(t+\Delta t) = \frac
	{{\bf r}_i(t)+\dot{\bf r}_i(t)\Delta t}
	{\lVert {\bf r}_i(t)+\dot{\bf r}_i(t)\Delta t \rVert}\,.
\end{equation}
We believe that this method, first proposed in~\cite{reza2006}, has been regarded as a standard procedure to obtain the numerical results in various  works such as~\cite{lohe2009,lohe2,zhu2013,tanaka2014,Ott2019,fplk,fplk2,lipton2021} where a numerical technique is not explicitly specified.

To know the local arc approximated by Eq.~\eqref{eq:cni} for small $\Delta t$, 
one can consider the rotation from ${\bf r}_i(t)$ to ${\bf r}_i^{\rm c}(t+\Delta t)$.
Let ${\bf a}_i(t,\Delta t)$ be the angle vector from ${\bf r}_i(t)$ to ${\bf r}_i^{\rm c}(t+\Delta t)$. Then, it follows that
${\bf r}_i(t) \times {\bf r}_i^{\rm c}(t+\Delta t)
=\hat{\bf a}_i \sin(\lVert {\bf a}_i \rVert)$, where ${\bf a}_i$ is 
the short notation for ${\bf a}_i(t,\Delta t)$ and $\hat{\bf a}_i$ is 
its direction vector. 
Since $\lVert {\bf a}_i(t,\Delta t) \rVert \propto \Delta t$
for small $\Delta t$, 
by $\sin(\lVert{\bf a}_i\rVert) = \lVert{\bf a}_i\rVert 
+O(\lVert{\bf a}_i\rVert^3)$, it follows that
${\bf a}_i = ({\bf r}_i(t) \times {\bf r}_i^{\rm c}(t+\Delta t))
(1+O((\Delta t)^2))$.
Using the rotation matrix $R({\bf a}_i)$, one may write 
${\bf r}_i^{\rm c}(t+\Delta t)=R({\bf a}_i){\bf r}_i(t)$.
Then, by substituting ${\bf r}_i^{\rm c}(t+\Delta t)$ in ${\bf a}_i\equiv
{\bf a}_i(t,\Delta t)$ with Eq.~\eqref{eq:cni}, one can obtain
\begin{equation}
	\label{cr2}
	{\bf r}_i^{\rm c}(t+\Delta t)
=R\left(({\bf r}_i(t)\times\dot{\bf r}_i(t))\Delta t(1+ O((\Delta t)^2))\right)
{\bf r}_i(t)\,,
\end{equation}
where the higher-order correction comes from the expansion of
$\lVert {\bf r}_i+\Delta t\dot{\bf r}_i \rVert$ for small $\Delta t$. 
Equation~\eqref{cr2} indicates that ${\bf r}_i^{\rm c}(t+\Delta t)$ lies on the great-circle tangent to $\dot{\bf r}_i(t)$ because
$({\bf r}_i\times\dot{\bf r}_i)\cdot {\bf r}_i=0$.

Using Eq.~\eqref{eq:2Kf}, one knows the rotation 
of the numerical method in Eq.~\eqref{eq:cni} 
takes place along 
\begin{equation}
\label{Ld}
{\bf r}_i\times \dot{\bf r}_i
={\bf v}_i-({\bf w}_i\cdot{\bf r}_i){\bf r}_i\,,
\end{equation} 
which is the angular velocity vector of that rotation.
This is different from ${\bf v}_i$, the angular velocity of 
the method in Eq.~\eqref{eq:hr} and, furthermore, that of the
model as explicitly written in Eq.~\eqref{eq:2Kf} that is equivalent to Eq.~\eqref{eq:2Kc}
in three dimensions.
The difference does not decrease
for any given $\Delta t$, 
Therefore, ${\bf r}_i^{\rm c}(t+\Delta t)$ in Eq.~\eqref{eq:cni} 
is not free from the artifact due to the incorrect rotation axis and
angular velocity.

The rotation of ${\bf r}_i^{\rm c}(t+\Delta t)$ becomes the
same as that of ${\bf r}_i^{\rm h}(t+\Delta t)$ 
when ${\bf w}_i=0$ for all $i$ (see Eq.~\eqref{Ld} that gives ${\bf v}_i$).
This explains why the numerical result of~\cite{reza2006},
where the method of Eq.~\eqref{eq:cni} was suggested,
is still valid: note ${\bf w}_i=0$ for all $i$ therein. 
However, this conventional numerical integration method 
${\bf r}_i^{\rm c}(t+\Delta t)$ in Eq.~\eqref{eq:cni}, which is 
considered as standard with almost no doubt on its validity, 
becomes improper as soon as non-vanishing ${\bf w}_i$ comes into play.
We emphasize again that the angular velocity with its own 
rotation axis of Eq.~\eqref{Ld} is not the correct one  
in Eqs.~\eqref{eq:2Kf} and \eqref{eq:hr}, and that
the difference remains $O(1)$ for any given $\Delta t$.

\subsection{\label{sn:cim} Ambiguity in rotations on sphere by a linear velocity}

\begin{figure}
\includegraphics[width=9.0cm]{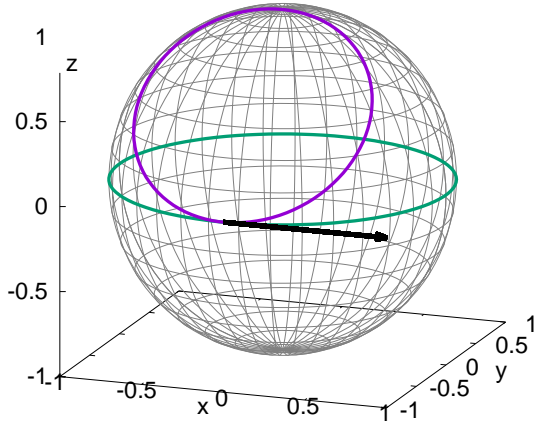}
\caption{(Color Online) Schematic figure explaining the ambiguity of rotation axis 
of the motion on a sphere for given linear velocity. 
The green circle represents the equator, and the purple one tangent 
to the equator is a circle that turns around from the arctic pole. 
The black arrow tangent to 
both circles represents the instantaneous velocity of a particle at the point 
where the circles meet. In this scenario, it is unclear whether the particle's 
movement should follow the purple or green circle. In fact, there are 
infinitely many circles tangent to the black arrow, along which the 
particle can move though we have only drawn two for simplicity.
}
\label{fig1}
\end{figure}

In this subsection, we analyze further where the problem of the conventional
method originates.
Figure~\ref{fig1} shows two circles that are tangent to a tangential vector 
at a point on sphere. 
Let the black arrow be the {\it linear} velocity $\dot{\bf r}_i(t)$ 
tangent to the sphere at ${\bf r}_i(t)$. 
The green circle 
is the great circle that is tangent to $\dot{\bf r}_i(t)$. 
The purple circle that turns around from the north pole is also tangent to 
$\dot{\bf r}_i(t)$.
Equation~\eqref{cr2} says 
${\bf r}_i^{\rm c}(t+\Delta t)$ 
lies on the great green circle.
Here, one may question if there is another circle 
that can be used for the position update with the condition that 
the difference rate approaches $\dot{\bf r}_i(t)$ 
in the $\Delta t\to 0$ limit. 
The purple circle shows the answer is yes 
because it is tangent to $\dot{\bf r}_i(t)$ at ${\bf r}_i(t)$.
When the position update is considered along the purple circle
with a displacement proportional to $\Delta t$ with a suitable coefficient, 
the difference rate can 
approach $\dot{\bf r}_i(t)$ in the $\Delta t\to 0$ limit.

Actually, there is an
infinite number of circles that are tangent to the black arrow
though only two are illustrated in Fig.~\ref{fig1}.
Therefore, a choice out of them in a way that is not justified by the model equation may result in an artifact. The method in Eq.~\eqref{eq:cni} is such an example.
The difference equation therein is not attributed to the model equation Eq.~\eqref{eq:2Kf} or, equivalently, Eq.~\eqref{eq:2Kc} in three dimensions.
Indeed, Eq.~\eqref{eq:2Kf} shows that the on-sphere motion of 
three-dimensional Kuramoto model is maintained not by a kind of a normalization to fix the off-sphere motion but by
a sequence of instantaneous circular motions on the sphere. 
Thus the update by Eq.~\eqref{eq:cni} is artificial, and the 
associated rotation axis and the angular velocity written in
${\bf v}_i-({\bf w}_i\cdot{\bf r}_i){\bf r}_i$ is no more than
the resultant artifact that is not curable 
for any numerical mesh $\Delta t$.

For an angular velocity ${\bf a}$,
the rotation property is encoded in $\Omega_{{\bf a}}$ (see Eq.~\eqref{eq:Om} for its definition). This is written in the 
rotation matrix 
$R({\bf a}\Delta t) \approx I+\Delta t\Omega_{\bf a}$ for small $\Delta t$~\cite{arfken,goldstein}, where $I$ 
is the identity matrix.  
The concept of $\Omega_{\bf a}$, 
already well-known for an infinitesimal transformation 
in traditional literatures, has led to the study of Lie algebra~\cite{wiki-it}.
The infinitesimal transformation of the conventional numerical method 
Eq.~\eqref{eq:cni} is represented by 
$\Omega_{{\bf v}_i-({\bf w}_i\cdot{\bf r}_i){\bf r}_i}$
while that for the method in Eq.~\eqref{eq:hr} is done 
by $\Omega_{{\bf v}_i}$.
Since  
this discrepancy does not disappear for any given $\Delta t$, 
the two numerical integration methods in Eqs.~\eqref{eq:cni} 
and \eqref{eq:hr} are distinct.
Thus, ${\bf r}_i^{\rm c}(t+\Delta t)$ [Eq.~\eqref{eq:cni}] 
is not a suitable method 
for the numerical integration of Eq.~\eqref{eq:2Kc} 
or, equivalently, Eq.~\eqref{eq:2Kf} in three dimensions.

Although the idea behind ${\bf r}_i^{\rm c}(t+\Delta t)$
seems to be valid, this has not been actually examined as above.
In short, 
the notion of Lie algebra realized with rotation axes as in the present study
is far beyond the simple procedure as in ${\bf r}_i^{\rm c}(t+\Delta t)$.
A direct integration with an accompanied normalization 
completely misses an on-sphere circular motion
specified by the rotation axis contained in the model equation.    
We were motivated to distinguish
${\bf r}_i^{\rm h}(t+\Delta t)$ of $\Omega_{{\bf v}_i}$ from 
${\bf r}_i^{\rm c}(t+\Delta t)$ 
by the observation that there are infinitely many circles 
tangent to each others at a point on sphere. 
The conventional ${\bf r}_i^{\rm c}(t+\Delta t)$ 
disregards the ambiguity caused by 
an infinite number of rotations corresponding to a given linear velocity.

\section{Comparison of the two methods}
\label{sn:cd}
In this section, it will be demonstrated that ${\bf r}_i^{\rm h}(t+\Delta t)$ 
reproduces known analytic results and expected behaviors of the oscillators 
while ${\bf r}_i^{\rm c}(t+\Delta t)$ does not.

\subsection{Simple analytic solution for one agent}
\label{bm}
The simple case that clearly shows that
${\bf r}_i^{\rm h}(t+\Delta t)$ [Eq.~\eqref{eq:hr}] 
yields correct behaviors, while  ${\bf r}_i^{\rm c}(t+\Delta t)$ [Eq.~\eqref{eq:cni}]
does not 
is when $N=1$ for constant ${\bf w}\neq 0$. 
In this case, from Eqs.~\eqref{eq:2Kf} and \eqref{eq:2Kc}, 
it reads that
\begin{equation}
	\dot {\bf{r}} = {\bf{w}}\times {\bf{r}} 
	= \Omega_{\bf w} {\bf{r}}\,,
\label{eq:od}
\end{equation}
where the subscript $i$ is omitted for simplicity.
With no loss of generality, 
we use ${\bf w}=(0,0,w)$ to write the solution as
\begin{equation}
	{\bf{r}}(t) = \left(\rho_0\cos(wt+\phi_0), 
	\rho_0\sin(wt+\phi_0), z_0\right)\,,
	\label{eq:sod}
\end{equation}
where $\rho_0=\sqrt{x_0^2+y_0^2}$ and $\phi_0=\arctan(y_0/x_0)$
for initial ${\bf r}(0)=(x_0,y_0,z_0)$ with $\lVert {\bf r}(0) \rVert=1$ 
(the case of $z_0=1$ that gives no motion is excluded).

For constant ${\bf w}=(0,0,w)$, one readily knows that 
the numerical solution of Eq.~\eqref{eq:od} by Eq.~\eqref{eq:hr} is
\begin{eqnarray}
\label{hsol}
{\bf r}^{\rm h}(n\Delta t)&=&R^n({\bf w}\Delta t){\bf r}(0)
\\
&=&\left(\rho_0\cos(wn\Delta t+\phi_0), 
\rho_0\sin(wn\Delta t+\phi_0), z_0\right) \nonumber
\end{eqnarray} for $n=1,2,...\,$. 
Then, all data points generated by ${\bf r}^{\rm h}(n\Delta t)$ lie on the solution circle in Eq.~\eqref{eq:sod},
and the occupation by the data points becomes denser
for smaller $\Delta t$. 
As a result, Eq.~\eqref{eq:hr} provides a proper 
numerical integration method for the analytic solution Eq.~\eqref{eq:sod}.
However, the method in Eq.~\eqref{eq:cni} does not, as shown below.

\begin{figure}
\includegraphics[width=9.0cm]{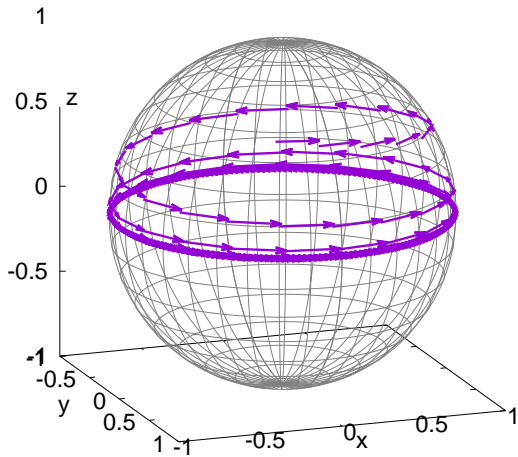}
\caption{(Color Online) Irreproducibility of the analytic solution Eq.~\eqref{eq:sod} by the conventional method Eq.~\eqref{eq:cni}.
The trajectory of Eq.~\eqref{eq:cni} unconditional approaches 
to the great circle
when $N=1$ and ${\bf w}\neq 0$ (see text for detail explanation). 
The great circle in this figure is the equator that 
is perpendicular to the used ${\bf w}=(0,0,2.2)$. 
Each arrow is the displacement during $2n\le t \le 2n+0.15$ for 
$n=0,1,2,...$. 
Though not shown, there is one round trip between adjacent arrows
(we did not display that because such details make 
visualization rather confusing). 
As the time-mesh, $\Delta t=0.01$ is used. 
}
\label{fig3}
\end{figure}

First of all, when Eq.~\eqref{eq:cni} is used, 
the first step location ${\bf r}^{\rm c}(\Delta t)=
\left({\rm r}(0)+\dot{\bf r}(0)\Delta t\right)
/\lVert \left({\rm r}(0)+\dot{\bf r}(0)\Delta t\right) \rVert$ 
is not on the solution circle in Eq.~\eqref{eq:sod},
unless $z_0 = 0$. This is because 
i) the $z$-component of ${\bf r}(0)+\dot{\bf r}(0)\Delta t$ is $z_0$ and 
ii) $\lVert {\bf r}(0)+\dot{\bf r}(0)\Delta t\lVert >1$.
The first comes from the fact that 
$\dot{\bf r}(0) = (0,0,w)\times{\bf r}(0)$ has no $z$-component, 
and the second comes from that ${\bf r}(0)$ of unit length 
is perpendicular to $\dot{\bf r}(0)$. It thus follows
that $z_0/(1+\delta_0) < z_0$ for $\delta_0\propto (\Delta t)^2 > 0$ ($z_0>0$ is assumed with no loss of generality). 
The decrease of $z$-coordinate value is repeated to show
$z_0>z_1>z_2>..$ with $z_{i+1}\equiv z_i/(1+\delta_i)$ for $\delta_i\propto (\Delta t)^2 > 0$.
Interestingly, the decrease at each step $\Delta_i\equiv z_i-z_{i+1}$ 
follows $\Delta_i \propto z_i$. 
Then, the updated data points 
approach to the equator at $z=0$ for small $\Delta t$,
as demonstrated in Fig.~\ref{fig3} with the numerical data
by Eq.~\eqref{eq:cni}.

The limit cycle rotating along the 
associated great circle
is always the long-term state of the numerical method Eq.~\eqref{eq:cni} independent of initial conditions when $N=1$ and ${\bf w}\neq 0$. This way, the conventional method
${\bf r}_i^{\rm c}(t+\Delta t)$ in Eq.~\eqref{eq:cni} cannot reproduce even 
the simplest analytic solution in Eq.~\eqref{eq:sod}.

\subsection{Collective long-term solutions with identical ${\bf w}_i$'s}
\label{csm}
\begin{figure}
\includegraphics[width=9.0cm]{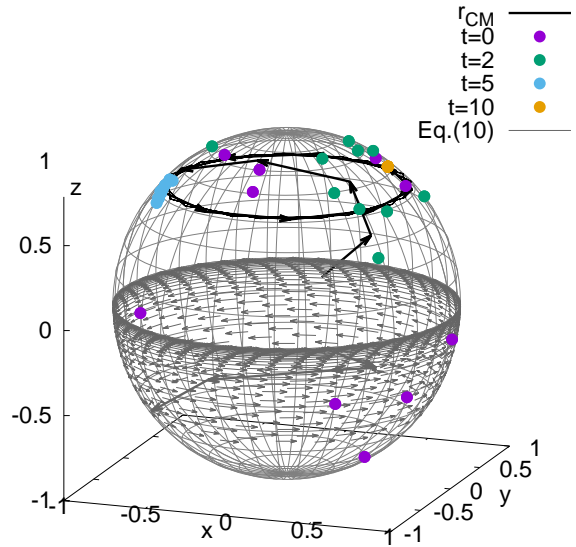}
\caption{(Color Online) [Upper hemisphere]
Formation of the limit cycle of collapsed 
agents in a system of size $N=10$  
by the method in Eq.~\eqref{eq:hr} when 
${\bf w}_i={\bf w}=(0,0,1)$ for all $i$. The solid circles 
with different colors represent the agents at the specified time, 
and the black bold 
arrow line is the displacement of the position center ${\bf r}_{\rm CM}$ per unit time.
Initially, the agents are randomly distributed on the surface. 
The sequence of the black bold line shows that the 
limit cycle is formed around $t=5$, and this is consistent with 
the analytic understanding in the text. 
We use $\Delta t = 0.01$ and $K=1$.
[Lower hemisphere] 
Result by the conventional method in Eq.~\eqref{eq:cni} for the same
numerical setting used above.
For the visualization and comparison, the data are displayed 
on the lower hemisphere by using $-z$ instead of $z$.
The four grey bold arrows are the displacement 
of ${\bf r}_{\rm CM}$ per unit time during $0\le t \le 4$. 
Each short grey arrow is the displacement
of ${\bf r}_{\rm CM}$ during $n\le t \le n+0.15$ for $n=4,5,..\,$. 
The final destination is the great circle 
perpendicular to $\bf w$, the equator when $\bf w$ is parallel to 
$z$-axis.
This observation on such great circle as the long-term state of Eq.~\eqref{eq:cni} is independent of the initial conditions
(see text for its reason). Note the limit cycle by Eq.~\eqref{eq:hr}
is not necessarily a great circle as illustrated in the upper hemisphere.
}
\label{fig4}
\end{figure}

For the system with ${\bf w}_i = {\bf w}$ for all $i$, using 
Eq.~\eqref{eq:hr}, we numerically observed that all agents converge to a 
rotating point of angular velocity ${\bf w}$ as time increases. 
We use ${\bf w}=(0,0,1)$ with no loss of generality. 
The results are illustrated on the upper hemisphere of Fig.~\ref{fig4}. 
All agents gather together before long; the data points at $t=10$ already 
look almost like a single point.
The deviation of all agents' locations is much smaller than the point size. 
Thereafter, ${\bf r}_i\approx {\bf r}_{\rm CM}$ so that 
${\bf k} \times {\bf r}_i\approx 0$. 
When these approximations are applied to Eq.~\eqref{eq:2Kf}, it follows that $\dot{\bf r}_{\rm CM} \approx {\bf w}\times {\bf r}_{\rm CM}$ 
whose structure is similar to that of Eq.~\eqref{eq:od}. 
Then, ${\bf r}_{\rm CM}$ readily gives the limit cycle like 
Eq.~\eqref{eq:sod}. The circle on a latitude in the upper hemisphere
of Fig.~\ref{fig4} is its numerical realization.

Analytically, if viewed in the rotating frame of angular velocity $\bf w$, 
the behavior of all particles/agents is simply a convergence to a fixed point by the positive coupling $K$. 
A compact mathematical proof of this behavior is provided in~\cite{zhu2013}. 
Thus, the collective long-term state is a limit cycle approaching 
a circle whose latitude is determined by the initial condition. 
The latitude of the limiting circle, which 
corresponds to the $z$-coordinate of the fixed point in the rotating 
frame, needs not to be zero but depending on initial conditions 
as mentioned.
However, this dependence is not the case for 
the numerical long-term state by Eq.~\eqref{eq:cni}.

The notable artifact of Eq.~\eqref{eq:cni} is that its long-term 
collective circular state is on the great circle at equator.
In the subsection~\ref{bm}, we examined the destination of a single agent with natural frequency vector parallel to $z$-axis is the equator at $z=0$ 
when Eq.~\eqref{eq:cni} is used for numerical integration.
Thus if there are more agents with the same natural frequency,
it is natural to expect that 
i) all trajectories will approach the equator at $z=0$ 
because this is the destination of each agent and 
ii) all agents will collapse to one point by the positive coupling $K$.
In the lower hemisphere of Fig.~\ref{fig4}, 
the grey arrows representing the motion of positions' center obtained by Eq.~\eqref{eq:cni}
correspond to the numerical observation given above.
In the repeated numerical tests, we observed that, 
regardless of the initial conditions, the long-term state is 
the rotation of collapsed data points along the great circle at $z=0$.
In general, the collective long-term state is 
the great circle whose plane is perpendicular to $\bf w$.
This is always the case for any numerical mesh $\Delta t$, 
and therefore the artifact of the conventional 
numerical method in Eq.~\eqref{eq:cni} is incurable.

A system of such agents with identical natural frequencies
was studied in~\cite{lipton2021}, and numerical data were 
presented. However, since the quantity in interest therein is 
invariant under the change in ${\bf w}$, the numerical data were 
obtained for ${\bf w}=0$. Our finding of
Eq.~\eqref{eq:hr} implies, if the ${\bf w}\neq 0$ case was 
numerically tested with Eq.~\eqref{eq:cni} or with 
one of its refinements but still neglecting  
$\Omega_{{\bf v}_i={\bf w}_i-{\bf k}\times{\bf r}_i}$, 
the result would become different. 
Meanwhile, we interestingly found in~\cite{zhu2013} 
such numerical trajectory that is similar to the one 
drawn in the upper hemisphere of Fig.~\ref{fig4}.
A simple way to obtain such data without using 
$\Omega_{{\bf w}_i-{\bf k}\times{\bf r}_i}$ is the following; first, prepare
the trajectory data of all agents 
using Eq.~\eqref{eq:cni} with ${\bf w}_i=0$ for all $i$,
and then rotate the data using ${\bf w}_i={\bf w}$
while considering the time label of each data point.
This is however merely a temporary substitute based on the proof 
in~\cite{zhu2013} for identical ${\bf w}_i$s, instead of a 
numerical method applicable to general situations.
For example, unless ${\bf w}_i$s are identical, 
such an attempt becomes groundless.

\subsection{Parallel ${\bf w}_i$'s: Apparent limit cycle and its disappearance}
\label{cm} 
The next case is that all ${\bf w}_i$s are not identical 
but parallel to each other, i.e., ${\bf w}_i \propto {\bf w}_j$ 
for all $i$ and $j$ (we still simply use ${\bf w}_i=(0,0,w_i)$). 
In this case, one may imagine a frustrating situation where the spreading 
of the agents along latitudinal direction
by the different $w_i$s contradicts the convergence 
by positive coupling $K$; 
the spreading prevents the convergence, and vice versa.
We note there is no such frustration along longitudinal direction. 
Thus, the agents
readily form an arc on a latitude as the longitudinal attraction 
by positive $K$ lasts without intervention.
Interestingly, there are two specific locations on the sphere, where 
the frustration does not happen.

The locations in interest are the poles, where 
the rotation axis (this is unique for all parallel ${\bf w}_i$s) 
penetrates the surface of the sphere.
The reason why the poles are special is 
that rotation at pole does not result in any 
actual motion. Thus, the conflict between spreading and converging disappears 
when all particles/agents move towards one of the poles. If this occurs, the system 
ultimately reaches a fixed point solution at the pole.

\begin{figure}
\includegraphics[width=9.0cm]{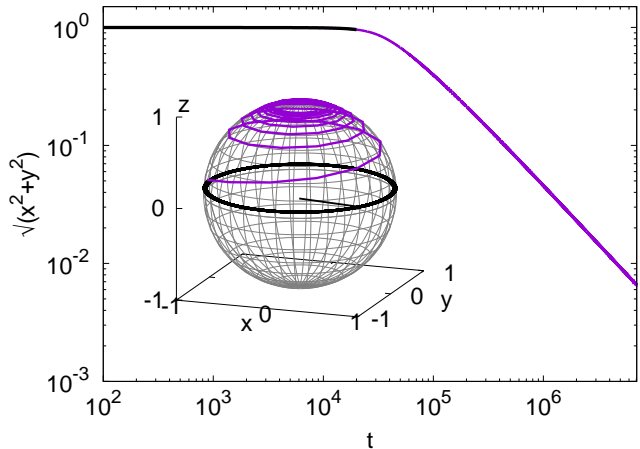}
\caption{(Color Online) Spiral to the fixed point by non-identical parallel ${\bf w}_i$s. The vertical value of the main plot is the numerical 
obtained distance between the position center $(x,y,z)$ 
at time $t$ of the agents in the system and the rotation axis; 
$z$-axis is chosen as the direction of the rotation, 
i.e., ${\bf w}_i = (0,0,\omega_i)$. As shown, 
$\sqrt{x^2+y^2}$ stays around $1$ even up to $t\approx 10^4$ 
(the horizontal black line). However, it decays thereafter 
following $\rho \sim t^{-c}$ for a positive $c$ 
(the declining purple line), where $c\approx 1$ by fitting the data. 
This asymptotic form is used in the inset to draw the spiral 
to the fixed point at the pole (the purple curve in the upper hemisphere).
The black curve in the inset shows the position center 
up to $t\approx 10^4$. We use $\Delta t= 0.01$, $K=32$, 
and $N=10$. Initially, the agents are randomly distributed. 
The $\omega_i$ are chosen from the uniform 
distribution with unit mean and the width $0.2$.}
\label{fig5}
\end{figure}

Figure~\ref{fig5}, drawn with data generated by Eq.~\eqref{eq:hr}, 
shows that the trajectory of the position center 
${\bf r}_{\rm CM}$ forms a spiral to the fixed point at a pole.
In the early stage, a limit cycle seems to appear, but eventually this approaches the fixed point at a pole in the end. Although a cycle lasts longer for larger $K$ and smaller $N$, it finally disappears in the numerical tests. The data obtained for $K=32$ and $N=10$ are shown in Fig.~\ref{fig5}. 
In the numerical simulations, we have not observed the other kind of long-term state except the fixed point at a pole 
for non-identical parallel ${\bf w}_i$s.

\begin{figure}
\includegraphics[width=7.5cm]{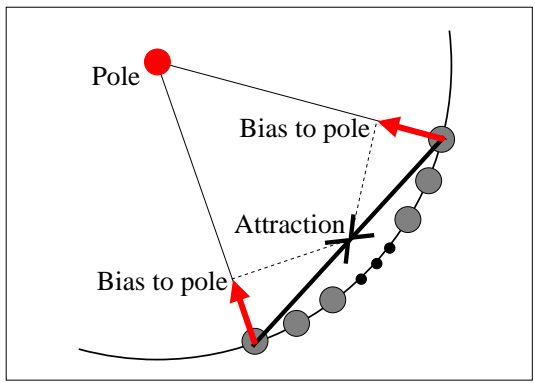}
\caption{(Color Online) 
Schematic diagram explaining the bias to the pole of agents that form an arc on a latitude. Here the geodesic path between the two end agents of the arc is represented by the straight line. The curve represents the latitude on which the arc of agents (grey circles) is formed, and the bold straight line is the geodesic path along which the attraction is applied. The attraction has such component that heads to the pole (see the red arrow). This is always the case whichever pair of agents is considered in the arc. Since the red arrows head to pole in all cases, their sum results in the bias to the pole of the arc.
}
\label{fig6}
\end{figure}

A possible explanation why ${\rm r}_{\rm CM}$ heads to the pole is following.
We first note that the spreading 
by the different $w_i$s occurs
in the lateral direction only. Since there is no disturbance in converging 
along longitudinal direction, the agents are expected to
form a cluster of arc on a latitude after an initial duration. 
The positive coupling causes each agent in a pair to attract the 
other along the geodesic direction between them.
The attraction on agent $i$ by agent $j$ is applied following 
$({\bf r}_i\times{\bf r}_j)\times{\bf r}_i$ [see Eq.~\eqref{eq:2K}], 
which is tangent to the geodesic curve between ${\bf r}_i$ and ${\bf r}_j$.
Here, we remark that the geodesic curve between any two points of the arc is on the northern side compared to the arc itself
(this is explained now for the upper hemisphere). 
See the diagram in Fig.~\ref{fig6}. 
As a result, the longitudinal component of any attraction always points 
towards the north pole.

Hence, all particles/agents are biased to the pole. The bias continues
as long as the agents form an arc with lateral spreading 
by different $w_i$s on a latitude.
The bias weakens as the arc approaches the pole. 
It vanishes when all agents converges the pole
to form the fixed point at the pole. 
If this process happens below the equator,
the final destination becomes the south pole.

The numerical result becomes different when the conventional method 
Eq.~\eqref{eq:cni} is used. For the same setting used above, 
${\bf r}_{\rm CM}$ approaches the equator, which looks like 
the long-term profile mentioned in subsections in~\ref{bm} and \ref{csm}
(not shown here). 
Numerically, it is observed that this long-term behavior persists until 
$K\approx 0.6$ and, below this value, ${\bf r}_{\rm CM}$ tends to approach the pole. 
We do not further study this seemingly transition because 
it is basically an artifact not belonging to the model equation 
Eq.~\eqref{eq:2Kf} or, equivalently, Eq.~\eqref{eq:2Kc}.

As far as we know, a limit cycle at a latitude
has been taken for granted when ${\bf w}_i$s are parallel for all $i$.
It has been argued in~\cite{lohe2009} that this is
a necessary condition for the existence of a limit cycle.
In contrast, no such limit-cycle has been observed in our numerical simulations. 
This does not logically matter at all
because a necessary condition does not guarantee 
the existence. In our numerical tests, the final feature is not a limit cycle 
but rather the fixed point at a pole, resulting from the direction of the 
parallel natural frequencies. 
This is consistent with our understanding of the pole as a fixed-point solution (see Fig.~\ref{fig6}).
Although an limit cycle persists for an extended period 
in some cases, but it eventually spirals towards 
the fixed point at the pole without exception in repeated numerical tests.

\section{\label{summary}Summary \& outlook}
In the present work, we have developed a new and improved
numerical integration method [Eq.~\eqref{eq:hr}] for the three-dimensional Kuramoto model.
It follows directly from the 
vector product form of the differential equation
as given in Eq.~\eqref{eq:2Kf}, which takes the form
of the rigid-body rotation on the unit sphere.
The associated numerical integration method then immediately follows 
if we use appropriate rotation matrices.
We have shown that the observation made in our paper that
the Kuramoto oscillators undergo rigid body rotations provides a useful perspective on 
the generalized Kuramoto model. 

\begin{figure}
\includegraphics[width=9.0cm]{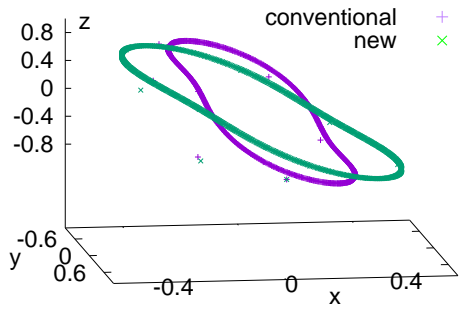}
\caption{(Color Online) 
Reproduction of Fig.~2 (a) in Ref.~\cite{tanaka2014} 
using the conventional method (the purple + data). 
Except a few scattered initial data, the data points almost form a steady loop that
appears to coincide with that in
Fig.~2 (a) of Ref.~\cite{tanaka2014}.
The green x data showing clear difference from the purple ones was obtained 
using our new method.
}
\label{figtc}
\end{figure}

On the other hand, we have demonstrated that the conventional 
numerical integration method 
[Eq.~\eqref{eq:cni}] contains an artificial bias which incorrectly constrains the motion
of agents
to the great circle 
in the long time limit in most cases in the present study.
This is caused by the lack of information on the instantaneous rotation axis
in the simple numerical integration scheme following the linear velocity
as given in Eq.~\eqref{eq:cni}.
When numerical results are presented 
in other works
on the three-dimensional Kuramoto model, 
we note that the associated numerical schemes are not usually specified.
We suspect that most of 
those numerical results are based on the conventional method,
and, since it is really a very simple straightforward procedure, it probably was taken for granted and has not been specified. 
In Fig.~\ref{figtc}, we have reproduced Fig.~2 (a) of Ref.~\cite{tanaka2014} using 
the conventional method (the purple + data). 
One can clearly see that the result in that paper resembles closely 
that obtained via the conventional method. 
As expected, our new method gives clearly different result (the green x data).
In light of the present work, we believe that
the previous numerical works on three-dimensional Kuramoto model need to be reexamined.

The improved numerical method we have presented is easy to implement 
once one identifies the corresponding rotation matrices. Yet, it is sophisticated enough 
to capture a subtle behavior of the oscillators in the long time limit in which
the apparent limit cycle produced by the conventional method eventually decays into a fixed point
as shown in Sec.~\ref{cm}. 
In order to apply the similar numerical scheme to higher dimensions,
a generalization of the Kuramoto model in the form of Eq.~\eqref{eq:hr} will be necessary, which is now under investigation.
The comparison of Eqs.~\eqref{eq:cni} and \eqref{eq:hr}
for more general values of the natural frequency vectors is also an interesting future project.

\section{Acknowledgments}
This research was supported by the NRF Grant No. 2018R1D1A1B07049254(H.K.L.) 
and 2021R1A2B5B01001951(H.H).

\def\tb{\textbackslash}
 
\end{document}